# A LONG-LIVED SHARP DISRUPTION ON THE LOWER CLOUDS OF VENUS

J. Peralta[*], T. Navarro[2,3,4], C. W. Vun[1], A. Sánchez-Lavega[5], K. McGouldrick[6], T. Horinouchi[7], T. Imamura[8], R. Hueso[5], J. P. Boyd[9], G. Schubert[2], T. Kouyama[10], T. Satoh[1], N. Iwagami[11], E. F. Young[12], M. A. Bullock[12], P. Machado[13], Y. J. Lee[14], S. S. Limaye[15], M. Nakamura[1], S. Tellmann[16], A. Wesley[17], and P. Miles[18]

[1]Institute of Space and Astronautical Science, JAXA, Sagamihara, Japan.
[2]Department of Earth, Planetary, and Space Sciences, University of California, Los Angeles, CA, USA.
[3]McGill Space Institute, McGill University, Montréal, QC, Canada.
[4]Department of Earth and Planetary Sciences, McGill University, Montréal, QC, Canada.
[5]Escuela de Ingeniería de Bilbao, UPV/EHU, Bilbao, Spain.
[6]Laboratory for Atmospheric and Space Physics, Univ. of Colorado Boulder, Boulder, CO, USA.
[7]Faculty of Environmental Earth Science, Hokkaido University, Hokkaido, Japan.
[8]Graduate School of Frontier Sciences, The University of Tokyo, Tokyo, Japan.
[9]CLaSP Department, University of Michigan College of Engineering, Ann Arbor, USA.
[10]National Institute of Advanced Industrial Science and Technology, Japan.
[11]Tokyo 156-0044, Japan.
[12]Southwest Research Institute, Boulder, CO 80302, USA.
[13]Institute of Astrophysics and Space Sciences, Portugal.
[14]Technische Universität Berlin, Berlin, Germany.
[15]Space Science and Engineering Center, University of Wisconsin, Madison, WI, USA.
[16]Rheinisches Institut für Umweltforschung, Universität zu Köln, Cologne, Germany.
[17]Astronomical Society of Australia, Murrumbateman, New South Wales, Australia.
[18]Gemeye Observatory, Rubyvale, Queensland, Australia.

May 15, 2020

## ABSTRACT

Planetary-scale waves are thought to play a role in powering the yet-unexplained atmospheric superrotation of Venus. Puzzlingly, while Kelvin, Rossby and stationary waves manifest at the upper clouds (65–70 km), no planetary-scale waves or stationary patterns have been reported in the intervening level of the lower clouds (48–55 km), although the latter are probably Lee waves. Using observations by the Akatsuki orbiter and ground-based telescopes, we show that the lower clouds follow a regular cycle punctuated between 30°N–40°S by a sharp discontinuity or disruption with potential implications to Venus's general circulation and thermal structure. This disruption exhibits a westward rotation period of $\sim$4.9 days faster than winds at this level ($\sim$6-day period), alters clouds' properties and aerosols, and remains coherent during weeks. Past observations reveal its recurrent nature since at least 1983, and numerical simulations show that a nonlinear Kelvin wave reproduces many of its properties.

***K*eywords** Venus · Planetary Atmospheres · Waves

---

[*]Corresponding author: J. Peralta (javier.peralta@ac.jaxa.jp).

**Plain Language Summary**

One of the biggest mysteries of Venus is its atmospheric superrotation that allows the atmosphere to rotate 60 times faster than the solid planet. Atmospheric waves are among one of the possible mechanisms thought to feed this superrotation by pushing energy to different locations of the atmosphere. In fact, the upper clouds of Venus located at 65–70 km exhibit varied giant waves, like the so-called Y-feature or the more recently-discovered bow-shape wave that keeps "stationary" over Aphrodite mountains. In contrast, these planetary-scale waves are missing at the deeper lower clouds (48–55 km). This is especially puzzling in the case of the stationary waves since the lower clouds are located between the upper clouds and the surface, where they are thought to be generated. Thanks to the high-quality observations of Venus from JAXA's space mission Akatsuki and NASA's IRTF telescope, we discovered at the lower clouds an intriguing sharp discontinuity that propagates to the west faster than the winds while altering the clouds' properties and suffering little distortions during weeks. A re-analysis of past observations revealed that this is a recurrent phenomenon that has gone unnoticed since at least the year 1983. Numerical simulations evidence that an atmospheric wave generated below the clouds and probably pumping energy to the upper clouds can explain many of its properties.

**Keypoints:**

- Discovery of an equatorial cloud discontinuity at the middle and lower clouds of Venus, where no planetary wave had been found before.

- This disruption propagates to the West faster than the winds, keeps coherent for weeks and alters clouds' properties and aerosols.

- Past observations confirm its existence since 1983. Numerical simulations suggest a physical origin as a nonlinear Kelvin wave.

# 1 Introduction

The atmosphere of Venus is dominated by superrotating winds that at the cloud layers (∼48–70 km) (Titov et al., 2018) exhibit speeds 60 times faster than the planet (Sánchez-Lavega et al., 2017). The clouds of Venus are mostly composed of $H_2SO_4$-$H_2O$ droplets and are stratified into three layers (Titov et al., 2018). In the upper clouds (56.5–70 km above the surface), absorbers of known and unknown composition are responsible for the dark markings observed in ultraviolet images and for most of the absorption of the solar energy not reflected by the clouds (Titov et al., 2018). The middle and lower clouds (hereafter, simply named *deeper clouds*) are within 47.5–56.5 km and importantly contribute to the greenhouse effect and the radiative energy balance. These clouds exhibit variable cloud optical thickness (Titov et al., 2018; McGouldrick et al., 2012; Peralta et al., 2019a) in a region where the lapse rate is close to adiabatic and convection dominates vertical transport, as shown by observations (Yakovlev et al., 1991; Hinson and Jenkins, 1995; Tellmann et al., 2009; Ando et al., 2020) and modelling (Imamura et al., 2014; Lefèvre et al., 2018). The middle clouds (50.5–56.5 km) are observed on the dayside at visible and near-infrared (900–1000 nm) wavelengths (Titov et al., 2018; Peralta et al., 2019a), while the lower clouds (47.5–50.5 km) are observed on the nightside using spectral windows at 1.74, 2.26 and 2.32 $\mu$m (McGouldrick et al., 2012; Titov et al., 2018; Limaye et al., 2018). Clouds' morphology and motions are different at each of these layers (Sánchez-Lavega et al., 2017; Peralta et al., 2019a; Titov et al., 2018; Limaye et al., 2018; Peralta et al., 2019b; Horinouchi et al., 2017).

It has long been proposed that planetary-scale waves could play a role in powering the superrotation (Sánchez-Lavega et al., 2017). Some of them manifest visually at the upper clouds (65–70 km), like the Y-feature (Peralta et al., 2015) and the stationary bow-shaped wave (Fukuhara et al., 2017). Some others manifest in the wind field like thermal tides (Kouyama et al., 2019), Kelvin and Rossby waves Imai et al. (2019) and stationary features (Peralta et al., 2017). However, no planetary-scale waves or stationary patterns have been reported in the intervening level of the lower clouds (Peralta et al., 2008, 2017, 2019b,a) (48–55 km), even though stationary waves are probably generated at the surface (Navarro et al., 2018).

## 2 Methods.

The nightside lower clouds of Venus were studied using 1,519 images acquired at 1.735, 2.26 and 2.32 $\mu$m during April–November 2016 by the IR2 camera (Satoh et al., 2017) onboard the Akatsuki orbiter (Nakamura et al., 2016), giving preference to 2.26-$\mu$m images (238 useful images), which were less affected by light contamination from the saturated dayside (Satoh et al., 2017). The periods January–February 2017, November–December 2018 and January 2019 were studied with 78 $K_{cont}$ (2.32-$\mu$m) images from the instruments SpeX and iSHELL (Rayner et al., 2003, 2012) at NASA's Infrared Telescope Facility (IRTF). We also reanalyzed 376 images from the Visible and Infrared Thermal Imaging Spectrometer (VIRTIS) covering April 2006 to October 2008 during the Venus Express (VEx) mission (Drossart et al., 2007). The dayside middle clouds were inspected from December 2015 to December 2016 using 984 Akatsuki/IR1 900-nm images (Iwagami et al., 2018), although October 2016 was covered with 29 1-$\mu$m images from a 0.5-m ground-based telescope since the phase angle from Akatsuki was large. The spatial resolution of Akatsuki images ranges 74–0.2 km/pixel depending on the distance between the spacecraft and the planet (Nakamura et al., 2016), while for ground-based observations it varies from 29–65 km (IRTF) to 400 km (0.5-m telescope) (Sánchez-Lavega et al., 2016). A summary of the imagery dataset is shown in Table S1.

### 2.1 Image Processing.

Akatsuki IR1 and IR2 cameras have CSD/CCD and PtSi-CSD/CCD detectors respectively, with dimensions 1024×1024 and their images present some specific problems (Satoh et al., 2017; Iwagami et al., 2018). IR2 images sensing Venus's nightside at 1.74, 2.26 and 2.32 $\mu$m in the calibration version of this work (v20180201) present a problem of light contamination with halation rings and a cross pattern that extends horizontally and vertically around the saturated dayside of the planet, spreading with multiple reflections along the PtSi detector (Satoh et al., 2017). IR2 images taken with the 2.26-$\mu$m filter were chosen for the characterization of clouds' morphology and motions, since the contamination is sufficiently reduced in them. We reduced the light contamination with an image processing procedure consisting on an adjustment of the brightness/contrast, followed by convolution with unsharp-mask image filter, and finally adaptive histogram equalization (Peralta et al., 2018, 2019b). This procedure is not totally efficient, and the effect of light contamination is yet apparent in some of the Figures in this work (Fig. 1). The images of the nightside acquired with the guide camera of IRTF/SpeX (Rayner et al., 2003) and a $K_{cont}$ filter were subtracted with sky images and flat-fielded corrected, though they lacked absolute calibration. Light contamination from the saturated dayside was efficiently reduced by subtracting these images with other acquired with a B$\gamma$ filter (2.18 $\mu$m). The processing technique afterwards was like the one applied on IR2 images, skipping the adaptive histogram equalization.

IR1 900-nm dayside images suffered from the added effect of smear noise, a brightness mismatch among the four quadrants of the camera sensor and a very small signal-to-noise (S/N) ratio (Iwagami et al., 2018). Except for the latter, these effects are reasonably well corrected in the calibration version v20180201. To increase the S/N of the IR1 images, we applied a photometric Minnaert correction (Peralta et al., 2019a) followed by brightness/contrast enhancement, smoothing with a radius of 2-3 pixels and a later unsharp-mask. This procedure successfully enhances the cloud features, especially in IR1 images with smaller phase angle (and better S/N), although it also enhances brightness mismatch among the quadrants of the sensor (see Figs. 1B, 3A and 3C). The images from small telescopes were acquired with a 508-mm Newtonian telescope, a FLIR GS3-U3-32S4M-C camera and a Thorlabs FELH1000 1-$\mu$m long-pass filter. These ground-based images covered 18 days from 9 to 30 of October 2016, with a solar elongation of 35°, a mean diameter of about 13 arcsecs, and an 80% of illuminated fraction. As with the IR1/900-nm images, we applied a Minnaert photometric correction followed by unsharp-mask (see examples in Figure S4).

### 2.2 Navigation of images.

Small uncertainties are known to affect the pointing of Akatsuki cameras, inhibiting high accuracy in the navigation of the Venus images (Ogohara et al., 2017). The image navigation was corrected using an ellipse fitting procedure where an automatic determination of the planetary limb pixels corrects the pointing (Ogohara et al., 2017; Satoh et al., 2017; Horinouchi et al., 2017). This automated method was used for the IR1 images but discarded in the case of IR2 and SpeX since light contamination and frequent darkening of the clouds' opacity make difficult the automatic identification of the planetary limb. For these images, we used a software tool (Peralta et al., 2018) which improves the visualization of the limb through image processing and allows to perform a visual adjustment of the position, size and orientation of the planet's grid. In the case of the IR2 images, the orientation of the navigation grid is kept unmodified, while the position of the grid was adjusted with a precision of 1/10 of a pixel. Images from SpeX, iSHELL and small telescope were firstly navigated using NASA's SPICE kernels, and both position and orientation of the grid were adjusted.

### 2.3 Calculation of cloud properties using VEx/VIRTIS images.

Due to the problem of light contamination in the images of Akatsuki/IR2, we considered only VEx/VIRTIS images (see Table S1) to study the effect of the cloud discontinuity on the optical depth and size parameter of the nightside lower clouds (see, Figs. 5B and 5C). Prior to calculating the optical thickness and size parameter, we performed a correction of the limb darkening in images at 1.74 and 2.30 $\mu$m, following the formula given by Wilson et al. 2008:

$$I_{1.74\mu m} = \frac{I'_{1.74\mu m}}{0.316 + 0.685 \cdot \cos EA} \quad (1)$$

$$I_{2.30\mu m} = \frac{I'_{2.30\mu m}}{0.232 + 0.768 \cdot \cos EA} \quad (2)$$

where $I'_{1.74\mu m}$ and $I'_{2.30\mu m}$ are the observed radiances and $EA$ is the emission angle.

The optical depth $\tau$ was calculated using VIRTIS images at 1.74 $\mu$m as $\tau = \log\left(I^{max}_{1.74\mu m}/I_{1.74\mu m}\right)$, where $I^{max}_{1.74\mu m}$ and $I_{1.74\mu m}$ are, respectively, the maximum value of radiance and the radiance at every pixel in the image. The size parameter $m$ was calculated following the method of Carlson et al. 1993 using the formula adapted for VIRTIS by Wilson et al. 2008: $m = (I_{1.74\mu m})/(I_{2.30\mu m})^{0.53}$, where $I_{1.74\mu m}$ and $I_{2.30\mu m}$ are the calibrated radiances of the Venus images at 1.74 and 2.30 $\mu$m.

### 2.4 Simulations with the IPSL Venus GCM.

The IPSL (Institut Pierre-Simon Laplace) GCM (General Circulation Model) is a full-physics model that includes, among other things, radiative transfer for solar and thermal radiations, a boundary layer scheme, topography, hybrid vertical coordinates, and a temperature-dependent heat capacity (Lebonnois et al., 2010, 2016; Garate-Lopez and Lebonnois, 2018). The resolution used for the longitude-latitude grid is 96×96, and the configuration is the same as in Garate-Lopez and Lebonnois 2018, who arbitrarily increased solar heating rates of the poorly constrained properties of the lower haze below the cloud base in order to match observed temperatures. Superrotation is fully developed in numerical simulations with this model after 300 Venus solar days of simulation.

The GCM predicts zonal speeds of 45 m s$^{-1}$ at the bottom of the cloud deck (48 km) (see Fig. 6D), slower than the 60 m s$^{-1}$ from in-situ measurements by Pioneer Venus descent probes (Counselman et al., 1980). To correct this, we added an extra term in the dynamic core of the model to the equation of zonal momentum for latitudes equatorward of 50°:

$$\frac{du}{dt} = \ldots + (u_f - u) \cdot k(P) \quad (3)$$

with $u$ being the zonal wind, $u_f$ the forced zonal wind profile, and $k(P)$ a pressure-dependent coefficient:

$$k(P) = 10^{-6} \cdot \frac{1 + \tanh\left(\log\left(P/10^5\right)\right)}{2} \quad (4)$$

The forced wind zonal profile $u_f$ is constructed from the temporally and zonally averaged zonal winds from the standard simulation of Garate-Lopez and Lebonnois 2018, increased by 30%. This relaxed simulation starts from the initial state of fully developed superrotation from the standard simulation, and it converges to increased steady averaged zonal winds in less than a Venus day.

## 3 Results.

To study the global opacity and morphology of the deeper clouds over several revolutions of the mean flow (∼5-6 day period; see Horinouchi et al. 2017; Peralta et al. 2018, 2019a), we constructed time composites combining equirectangular projections of the images shifted according to the zonal background wind (see Figure 1; animations of panels 1C and 1E can be found in animated Fig. 2). At low latitudes, the nightside lower clouds show a variable dark band (higher opacity) (Crisp et al., 1991), while bright (lower opacity) bands dominate at mid-latitudes. The lower clouds drift to the west following a ∼5-6-day cycle. The equatorial band dominated by dark featureless clouds (longitude drift 40°–100° in Fig. 1A or 240°–300° in Fig. 1B) is observed to become narrower as mid-latitude bands gradually invade lower latitudes. Simultaneously, it exhibits bright swirls and other patterns reminiscent of von Kármán vortex streets (Fig. 1A) and the borders with the mid-latitude bands adopt a wavy shape (wavelengths ranging 4,000–6,000 km) with mesoscale billows and vortices (Peralta et al., 2019b; Horinouchi et al., 2017; Satoh et al., 2017). The cycle ends when the dark band develops brighter clouds that are abruptly interrupted by a sharp discontinuity or disruption. During August 2016 (Fig. 1A), an equatorial jet was observed (Horinouchi et al., 2017) ∼150°-200° west from the disruption. In October 2016 (Fig. 1C), the discontinuity became weaker, the equatorial jet seemed missing and the mid-latitude bands merged at the equator forming a bright trough (Peralta et al., 2019b). An examination of 85 radio-occultation profiles within 30°N-30°S obtained during 2006–2016 by VEx and Akatsuki reveals that this cycle implies night-time variations of 2–6% in atmospheric temperature, pressure and molecular density (see Fig. S1).

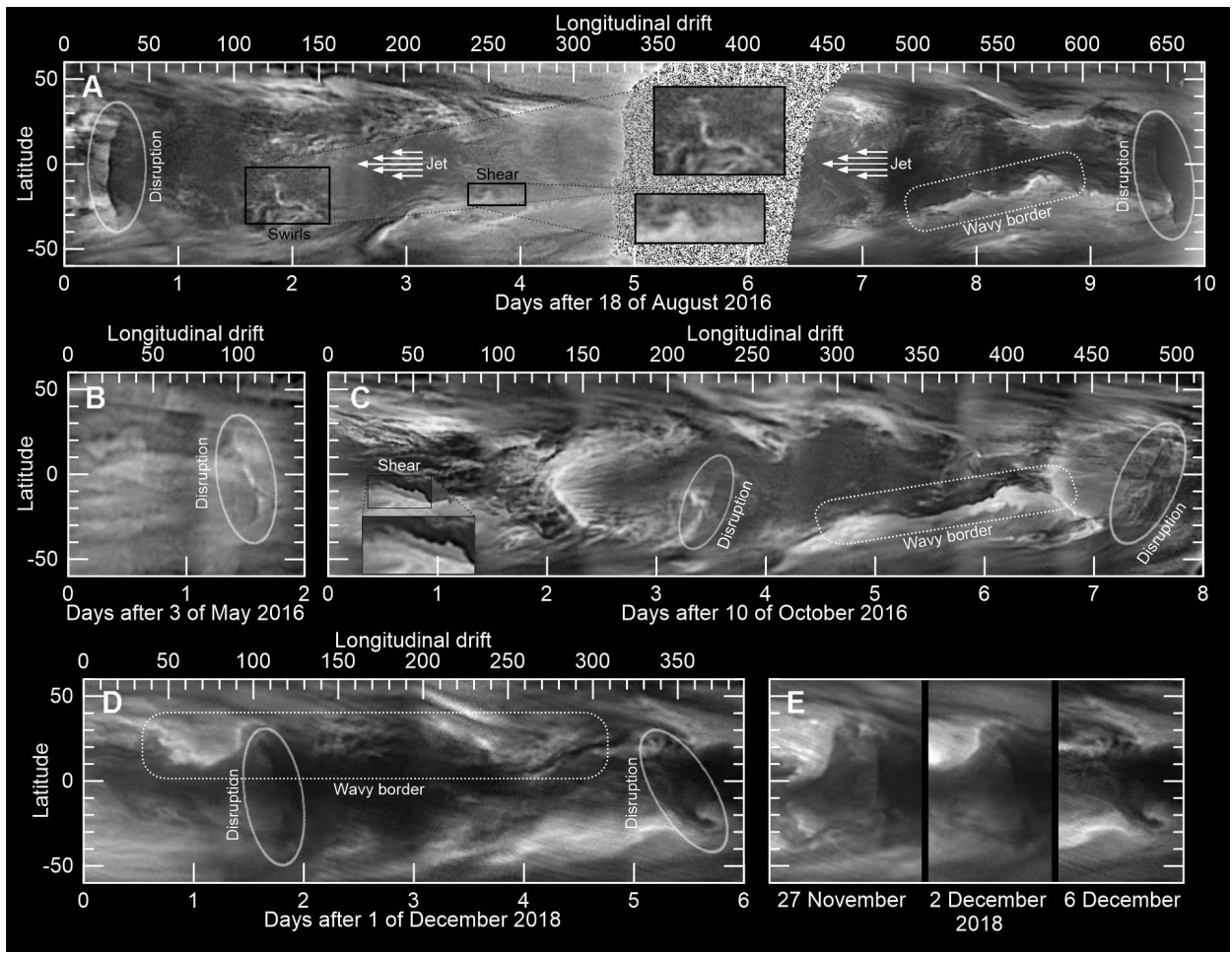

Figure 1: **The cycle of the deeper clouds of Venus.** Time composites of the deeper clouds displayed with overlapped equirectangular projections (60°N–60°S and 0.5° per pixel) placed from left to right with increasing dates: **(A)** 2016 August 18–22 and 25–27 (IR2-2.26 $\mu$m), **(B)** 2016 May 3–4 (IR1-900 nm), **(C)** 2016 October 10–17 (IR2-2.26 $\mu$m) and **(D)** 2018 December 1–6 (SpeX-2.3 $\mu$m). Panel **(E)** exhibits the evolution of a case of disruption in November 27 and December 2 and 6 of 2018 (SpeX-2.3 $\mu$m). All images display the nightside of Venus except **(B)**.

The sharp cloud discontinuity is a recurrent phenomenon at both lower and middle clouds (Figs. 1B, 3A and animated Fig. 4), although it is apparently missing in observations of the upper clouds (Yamazaki et al., 2018; Satoh et al., 2017) (Fig. 3C). The discontinuity is sometimes followed by undulations with wavelengths of 65±14 km (Fig. 3B), and can also extend thousands of kilometers westward from their northernmost end giving birth to sharp dark stripes (Figs. 1A and 1D) previously reported (Peralta et al., 2019b). Table S1 contains a summary of all the events of cloud disruptions. During 2016, 35 events were identified in the Akatsuki observations (Figs. S2–S3), and probably two more in observations with small telescopes (Figs. S4A–B). IRTF/SpeX observations between 2017 and 2019 revealed 7 events (Figs. S4C–D). A reanalysis of published ground-based observations (Allen and Crawford, 1984; Crisp et al., 1991; Bailey, 2006; Peralta et al., 2018) shows that the disruption was present on Venus's lower clouds in September 1983, January-February 1990, December 2005, July 2012 (Fig. 3D), and at least 12 times during 2006–2008 in VEx/VIRTIS images (see Figs. S4E–H).

Figure 2: **Animated Time Composites of the lower clouds of Venus during 10–16 of October 2016 and 1–10 of December 2018.** This animated figure displays satellite projections of the time series of the nightside lower clouds of Venus shown in Figures 1C and 1D in the main article. The first time series was constructed with 2.26-$\mu$m images obtained by the IR2 camera onboard JAXA's orbiter, and compress dates from 10 to 16 of October 2016. The second time series was made with images from ground-based observations by the instrument SpeX at NASA's IRTF infrared telescope. Before being combined to construct the time series, all the individual images were projected onto equirectangular geometry between 60°N–60°S and with a resolution of 0.5° per pixel. The satellite projections are centred at 0° latitude (equator) and 00:00 local time (midnight).

Figure 5A displays the rotation period, orientation and latitudinal extent of the disruption from 2016 until early 2019. The disruption appears within latitudes 30°N–40°S, it can have a length that varies from 800±50 to 7,600±200 km and has a mean width of 280±140 km (cross-to-along ratio ∼1:13). Its mean orientation relative to the equatorial plane is 85°±18° but ranges from 35° to 132°. In rare cases, the cloud discontinuity is seen split into two or three elements with different orientations (Fig. 3C). During August 2016, the disruption kept approximately coherent for ∼20 days. However, in general, the disruption suffers distinguishable changes after one revolution (Figs. 3A and S2), and its morphology and hemispherical symmetry/asymmetry seem unrelated to surface elevations (see Fig. S5). The disruption propagates to the west with a mean zonal speed of $-91 \pm 9$ m s$^{-1}$ –similar to the equatorial jet when this is present (Horinouchi et al., 2017)– and faster than the background winds within 30°N–40°S (see Fig. 1E), which are $-68 \pm 9$ m s$^{-1}$ at the lower clouds (48–55 km) (Peralta et al., 2018) and $-74 \pm 9$ m s$^{-1}$ at the middle clouds (∼55–65 km) (Sánchez-Lavega et al., 2017; Peralta et al., 2019a). Considering the overall data, the rotation period of this disruption is 4.9±0.5 days (Fig. 5A). Separate analyses of data for the middle and lower clouds yield periods of 4.7±0.4 and 5.0±0.5 days, respectively. The zonal drift of the disruption experienced larger variations in September (10 m s$^{-1}$) and November 2016 (∼30 m s$^{-1}$). Observations during November 2018 and January 2019 suggest that the drift of this feature increased by 40 m s$^{-1}$ during its propagation on the dayside (Figs. 5A and 1E).

VEx/VIRTIS images were used to study other effects of the disruption. On the nightside, the passage of the discontinuity implies radiance decreases of 75±7%, 88±7% and 88±6% at 1.74, 2.26 and 2.32 $\mu$m respectively (see Table S2), while on dayside IR1/900-nm images it implies albedo changes of only 1%–4% (Peralta et al., 2019a). The optical thickness (Fig. 5B) is observed to increase one order of magnitude west-to-east across the disruption, while its effect over the size parameter (regarded to be a proxy for particle size; see Carlson et al. 1993) is more variable although the west-side of the disruption is frequently linked to more abundance of smaller particles (Fig. 5C). This scenario seems consistent with downwelling and clouds' evaporation (McGouldrick et al., 2012) (hence, lower optical thickness) west of the disruption, and upwelling accompanied by formation of clouds (larger optical thickness) on its east-side. This upwelling combined with the increased $H_2SO_4$ vapor pressure and the larger nucleation rates expected for the high $H_2SO_4$ concentrations of the clouds (Sihto et al., 2009; Titov et al., 2018) can help the cloud condensation nuclei (CNN) of submicron size to overcome the Kelvin barrier (i.e. greater saturation pressure over smaller particles) and grow to small droplets with radii of $\sim 1$ $\mu$m (Imamura and Hashimoto, 2001), probably explaining the abundance of smaller particles observed in Fig. 5C.

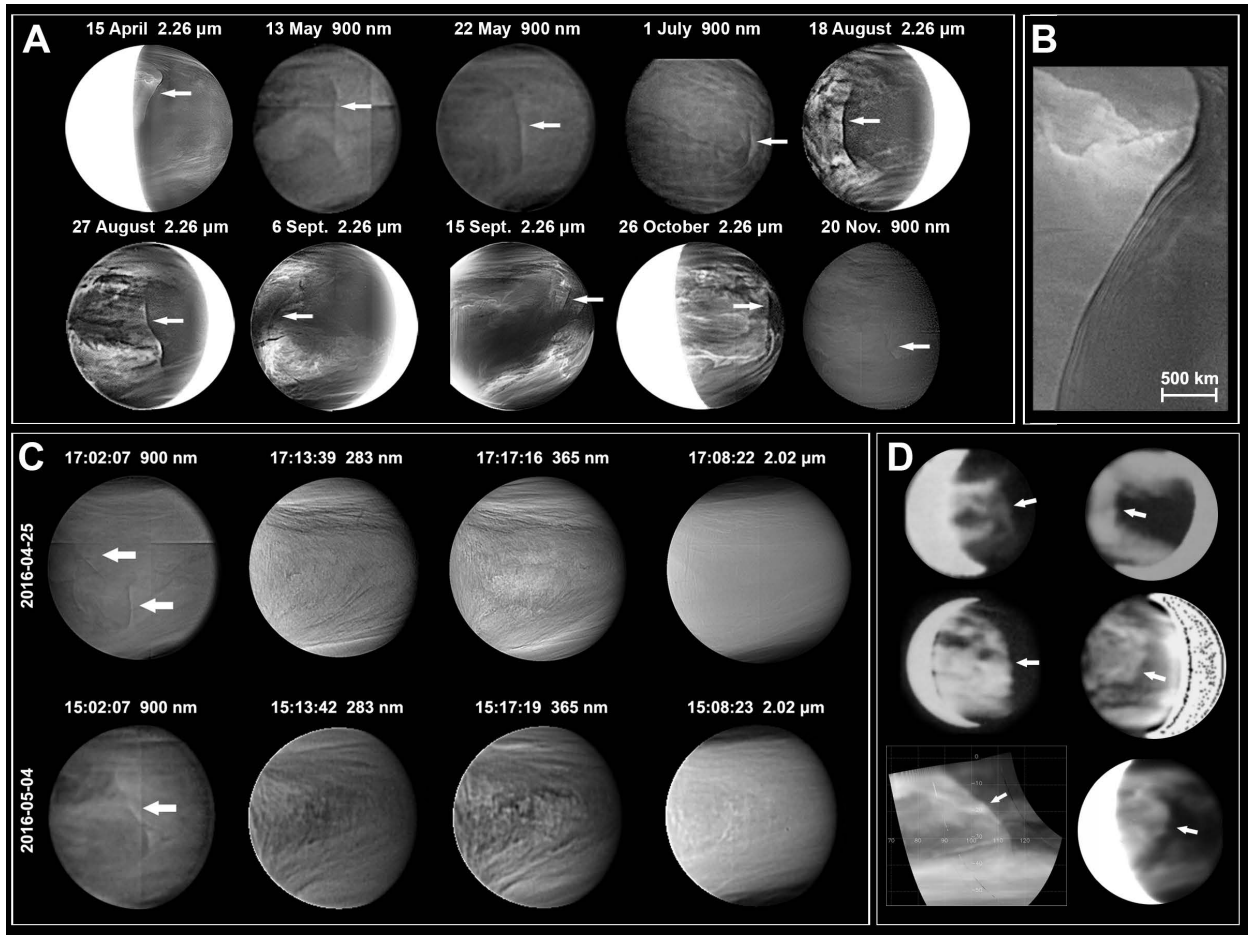

Figure 3: **Morphology, vertical extension and recurrence of Venus's disruption.** **(A)** morphological changes of the cloud discontinuity during 2016 on images from IR1/900-nm (dayside) and IR2/2.26-$\mu$m (nightside). Full set in Figs. S2–S3; **(B)** example of undulations behind a discontinuity in 15 April 2016; **(C)** discontinuities apparent on the middle clouds (IR1) but not at the upper clouds sensed with ultraviolet (UVI) and 2.02-$\mu$m images (IR2); **(D)** past events of disruption (left-to-right, top-to-bottom): September 1983 (Allen and Crawford, 1984), January–February 1990 (Crisp et al., 1991), December 2005 (Bailey, 2006), September 2006 (VEx/VIRTIS) and July 2012 (Peralta et al., 2018). All images were high-pass filtered (see subsection 2.1). Disruptions are marked with arrows.

Considering its morphology, long-term coherence and its westward drift faster than the mean flow, the disruption might be the manifestation of a weakly-dispersive Kelvin front (a nonlinear Kelvin wave). Kelvin fronts can be often undular like observed in Figure 3B, exhibiting gravity wave resonances excited behind the leading edge and propagating with the same phase speed (Fedorov and Melville, 2000). Day-night differences in stability and wind shear can change the intrinsic phase speed of a Kelvin wave (Peralta et al., 2015) and explain its faster propagation during its dayside passage (Fig. 5A). Simulations of the deeper clouds with the IPSL Venus GCM (Garate-Lopez and Lebonnois, 2018) (which does not incorporate interactions between circulation and clouds' processes) show that Kelvin fronts arise between 45–68 km under realistic atmospheric conditions (Scarica et al., 2019), they affect zonal and meridional speeds and –more weakly– temperatures (Fig. 6), and they barely interact with mountain waves

Figure 4: **The lower clouds of Venus during 15–30 of August 2016.** This animated figure displays two full cycles of the nightside lower clouds of Venus as observed at 2.26 $\mu$m by the IR2 camera onboard JAXA's orbiter Akatsuki. The passage of the equatorial cloud disruption can be observed during August 18–19 and 27–28. The images were processed to highlight finer cloud patterns (see sections Methods in main article) and projected onto a satellite geometry centered at 0° latitude (equator) and 00:00 local time (midnight). Latitudes 60°N, 30°N, 0°, 30°S and 60°S are displayed with dotted lines. Although the image processing highly reduces the problem of the light contamination in the IR2 images Satoh et al. (2017), its effect is yet apparent in this animation.

(Fig. S6). At 55 km, convection dominates vertical motions, explaining why the Kelvin wave is not apparent in Fig. 6G. The sharpest gradient/discontinuity in zonal winds is found at 57 km (Fig. 6B) and periodograms at different altitudes (Fig. 6H) evidence that the wave is trapped within the deeper clouds with a westward rotation period of ∼5.7 days, slower than the 4.9-day average from observations but between minimum/maximum periods reported. The rotation period is sensitive to the zonal winds below the clouds, being closer to observations when setting the relaxed profile of winds in the GCM (see subsection 2.4).

## 4 Conclusions.

While stationary waves seem abundant at the upper clouds of Venus (56.5–70 km), the middle and lower clouds (47.5–56.5 km) do not exhibit stationary waves but a dark band with cyclical behavior and a sharp cloud discontinuity at equatorial latitudes with long-term coherence and apparently unrelated to the Venusian topography. The absence of observable signatures of this discontinuity at the upper clouds, where waves exhibit an evanescent nature (Imai et al., 2019) contrasts with its quasi-permanent nature at the middle and lower clouds. Simulations with the IPSL Venus GCM show that a nonlinear Kelvin wave generated below the clouds (Yamamoto and Tanaka, 1997) reasonably reproduces this disruption and many of its observed properties (Fig. 6). Bore waves generated by katabatic fronts at the surface (Magdalena et al., 2006) or at higher altitudes by convective entrainment (Haghi et al., 2017) are alternative explanations to be explored by future studies, which might elucidate the role of this feature in the transport of atmospheric momentum and aerosols in the clouds of Venus.

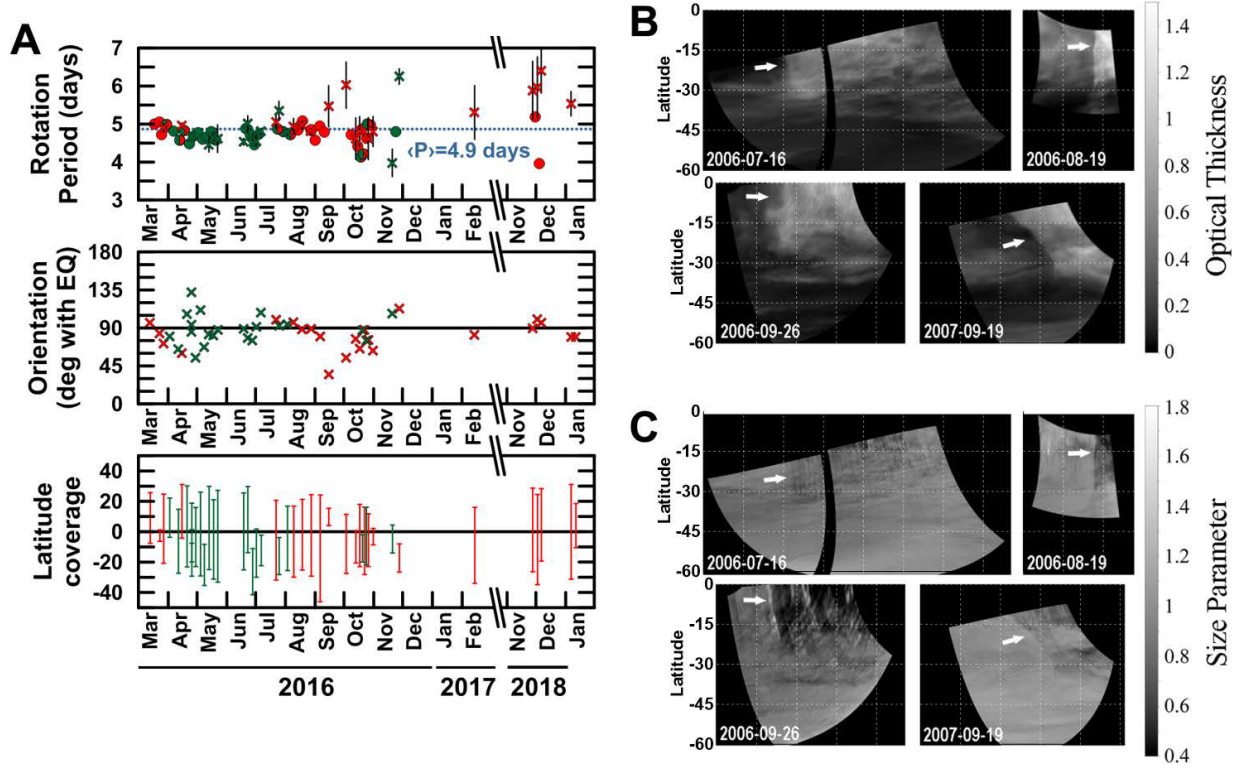

Figure 5: **Properties of the cloud disruption.** (A) Rotation period (mean period with blue-dotted line), orientation and latitude coverage of the disruption along 2016–2019. The periods were measured from the position of the disruption at the equator in images separated by hours (crosses) and several days (dots). When the disruption did not intersect the equator, we considered its longitude closest to the equator. Day/nightside data are shown in green/red, respectively. (B)–(C) Effect of the disruptions over the optical thickness (1.74 $\mu$m) and size parameter (1.74 and 2.32 $\mu$m) in equirectangular projections (0°–60°S, 0.2°·pix$^{-1}$) of VEx/VIRTIS images (see subsection 2.3).

## Acknowledgements

The dataset of Akatsuki is available at the public repository of JAXA (http://darts.isas.jaxa.jp/planet/project/akatsuki/). IRTF/SpeX images can be requested on reasonable request and will become available at the NASA/IPAC Infrared Science Archive (https://irsa.ipac.caltech.edu/applications/irtf/). VEx dataset is available at ESA's public repository (ftp://psa.esac.esa.int/pub/mirror/VENUS-EXPRESS/). Data S1 includes the IPSL Venus GCM simulations of Figure 6 and they can be downloaded from Zenodo (https://zenodo.org/record/3817479).

J.P. acknowledges JAXA's International Top Young Fellowship. T.N. and G.S. thank NASA's Grant NNX16AC84G. A.S.-L. and R.H. were supported by Spanish project AYA2015-65041-P (MINECO/FEDER, UE) and Grupos Gobierno Vasco IT-765-13. N.I. thanks partial support by JSPS KAKENHI Grant JP16H02225. Y.J.L. received funding from EU H2020 MSCA-IF No.841432. P.M. acknowledges FCT's project P-TUGA PTDC/FIS-AST/29942/2017. S.S.L. thanks NASA's Grant NNX16AC79G. All authors acknowledge the hard work done by the Akatsuki team. Sincere thanks to James O'Donoghue for proofreading this paper.

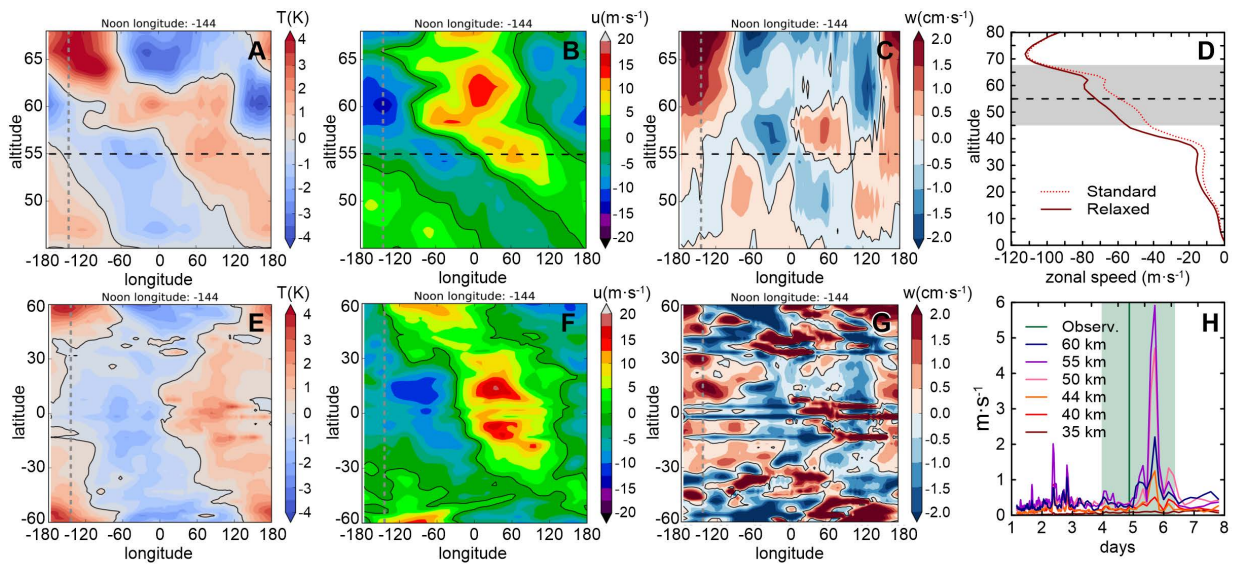

Figure 6: **Kelvin wave according to a Venus GCM.** Wave disturbances on temperatures and zonal and vertical speeds are shown as vertical cross-sections within 45–68 km **(A–C)** and horizontal ones at 55 km **(E–G)**. **(D)** shows the standard and relaxed profiles of zonal winds (see subsection 2.4), with the grey-shaded area marking the altitudes in **(A–C)**. **(H)** displays periodograms of the zonal speeds at several altitudes, with the green line/shaded-area standing for the averaged and minimum/maximum rotation periods from observations of the disruption. Noon longitude and 55-km altitude are shown with grey and black dashed lines, respectively.

**Supplementary Figure S1**

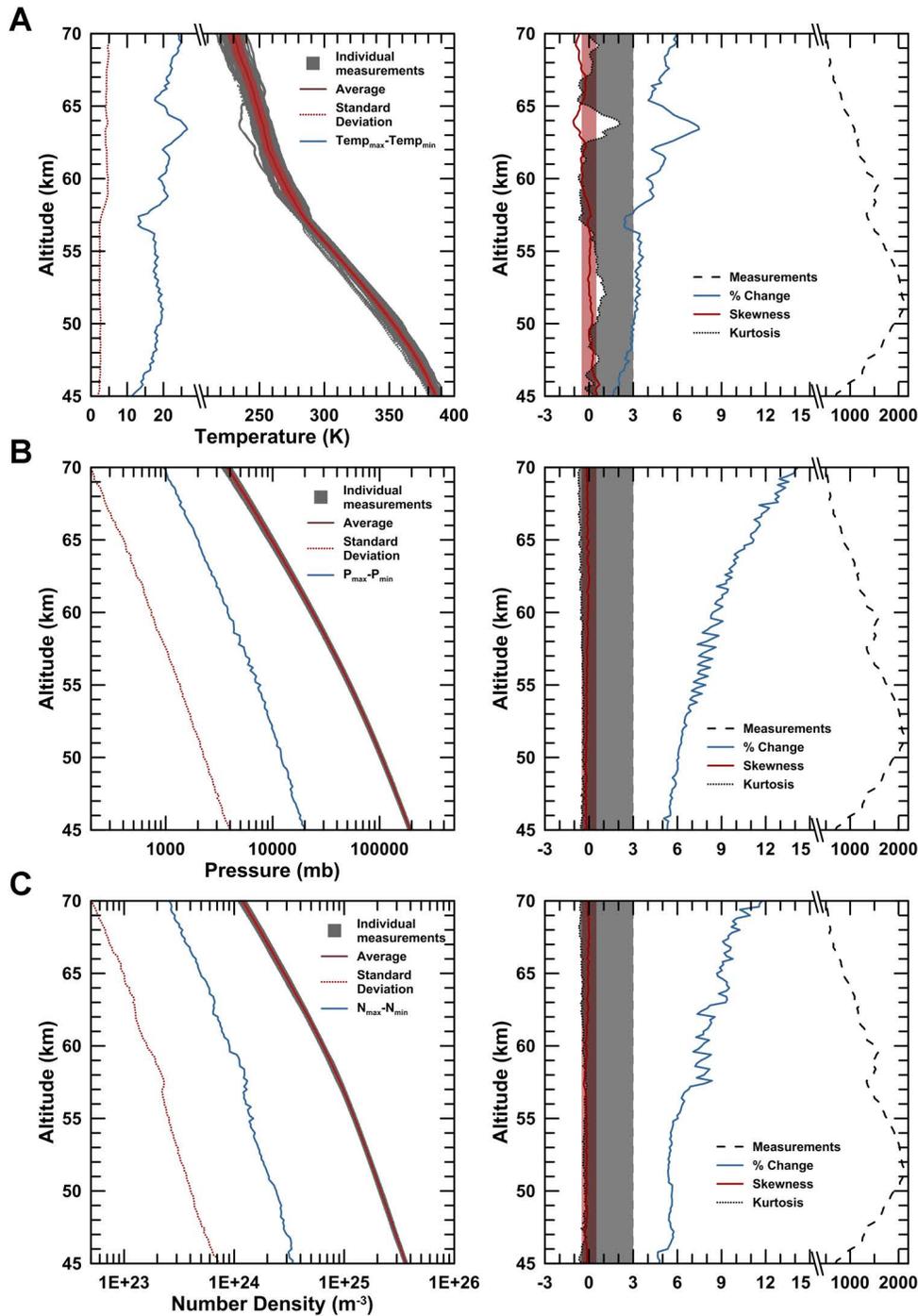

Figure 1: **Atmospheric parameters on the night of Venus from radio-occultation measurements.** Vertical profiles of the night temperature (A), pressure (B) and number density (C) from 85 radio-occultation profiles within 30°N–30°S and during 2006–2016 by Venus Express and Akatsuki. Vertical bins of 200 m were used for calculations. In right column: percent of change for the maximum deviation (relative to the mean), Skewness, Kurtosis and number of measurements within each vertical bin. The dark red area stands for symmetric data limited by Skewness values -0.5 and +0.5. The grey area stands for Kurtosis lower than 3 (no heavy tails or outliers).

**Supplementary Figure S2.**

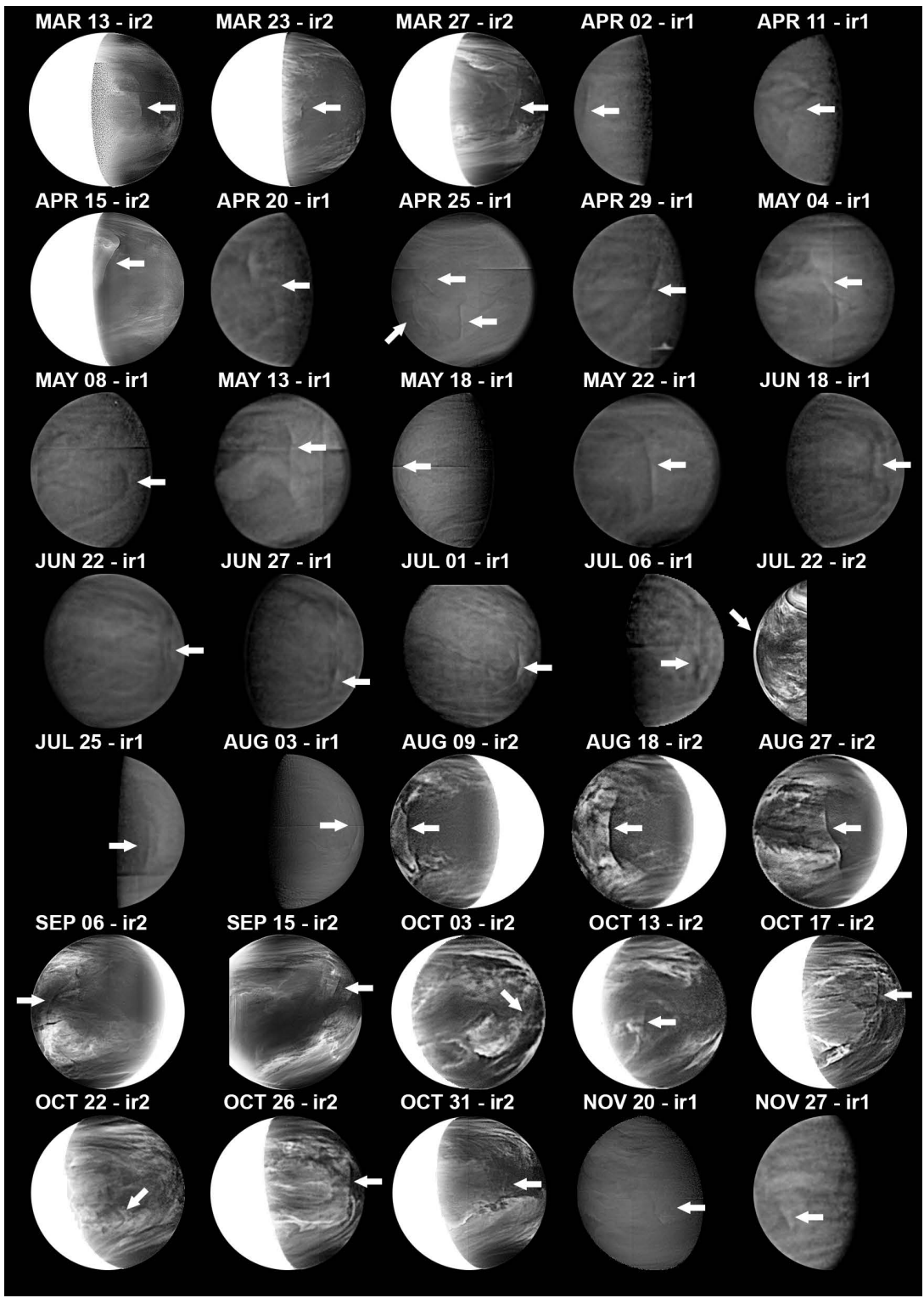

Figure 2: **Akatsuki observations of the clouds' disruption during 2016.** IR2/2.26-$\mu$m (nightside clouds' opacity) and IR1/900-nm images (dayside albedo) are displayed. The images were processed as described in the main work. Events of the disruption were identified in the dates of the year 2016 exhibited in the Figure.

**Supplementary Figure S3.**

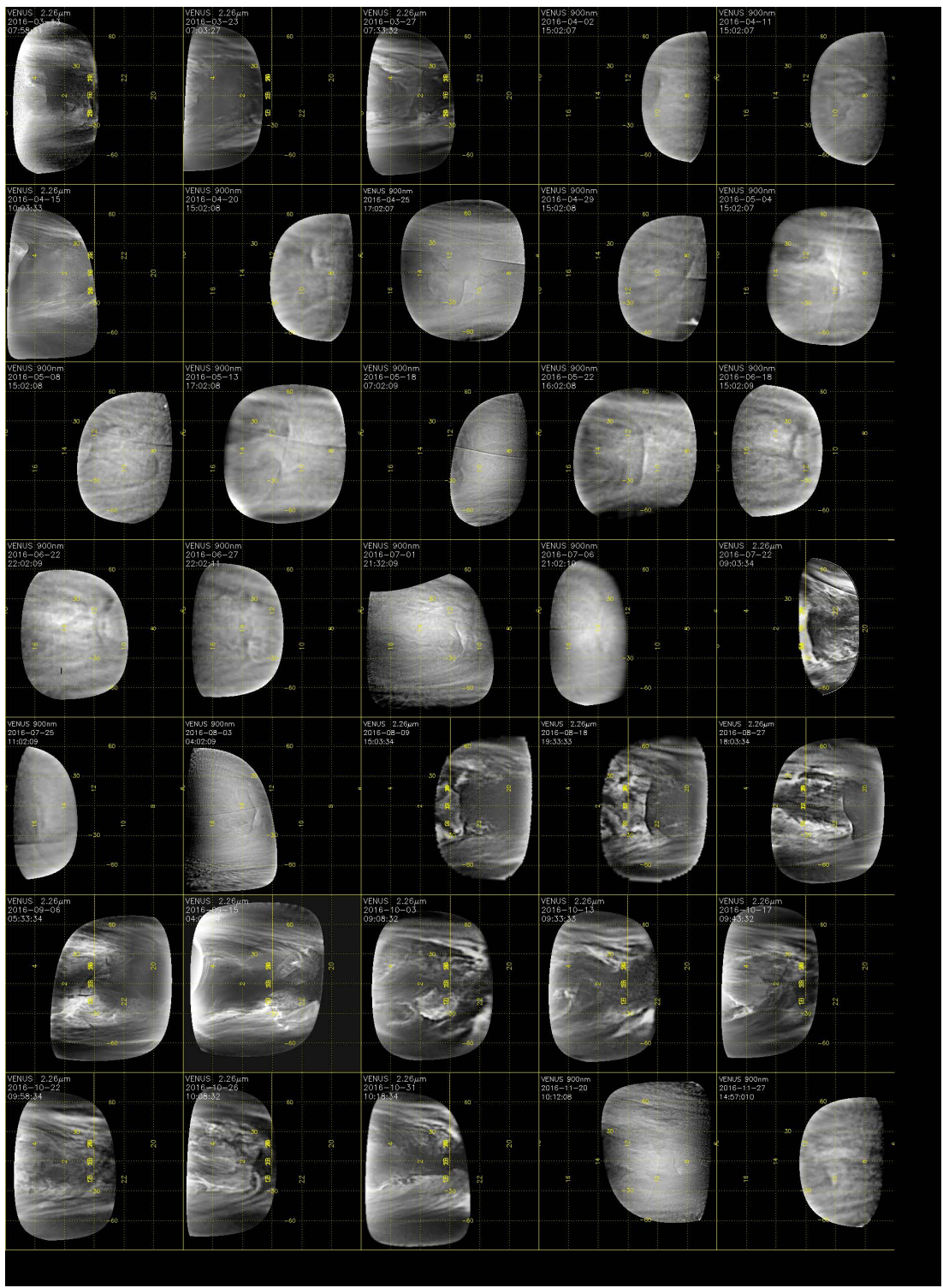

Figure 3: **Cylindrical projections of Akatsuki observations of the cloud disruption during 2016.** Contents are the same as in Extended Figure 1 although with the Akatsuki images displayed as equirectangular projections with grid resolution of 0.5° per pixel and latitude and local time boundaries 90°N–90°S and 06h–18h (dayside) and 18h–06h (nightside). The dates and times for each image are displayed at the left-up corner or each projection.

**Supplementary Figure S4.**

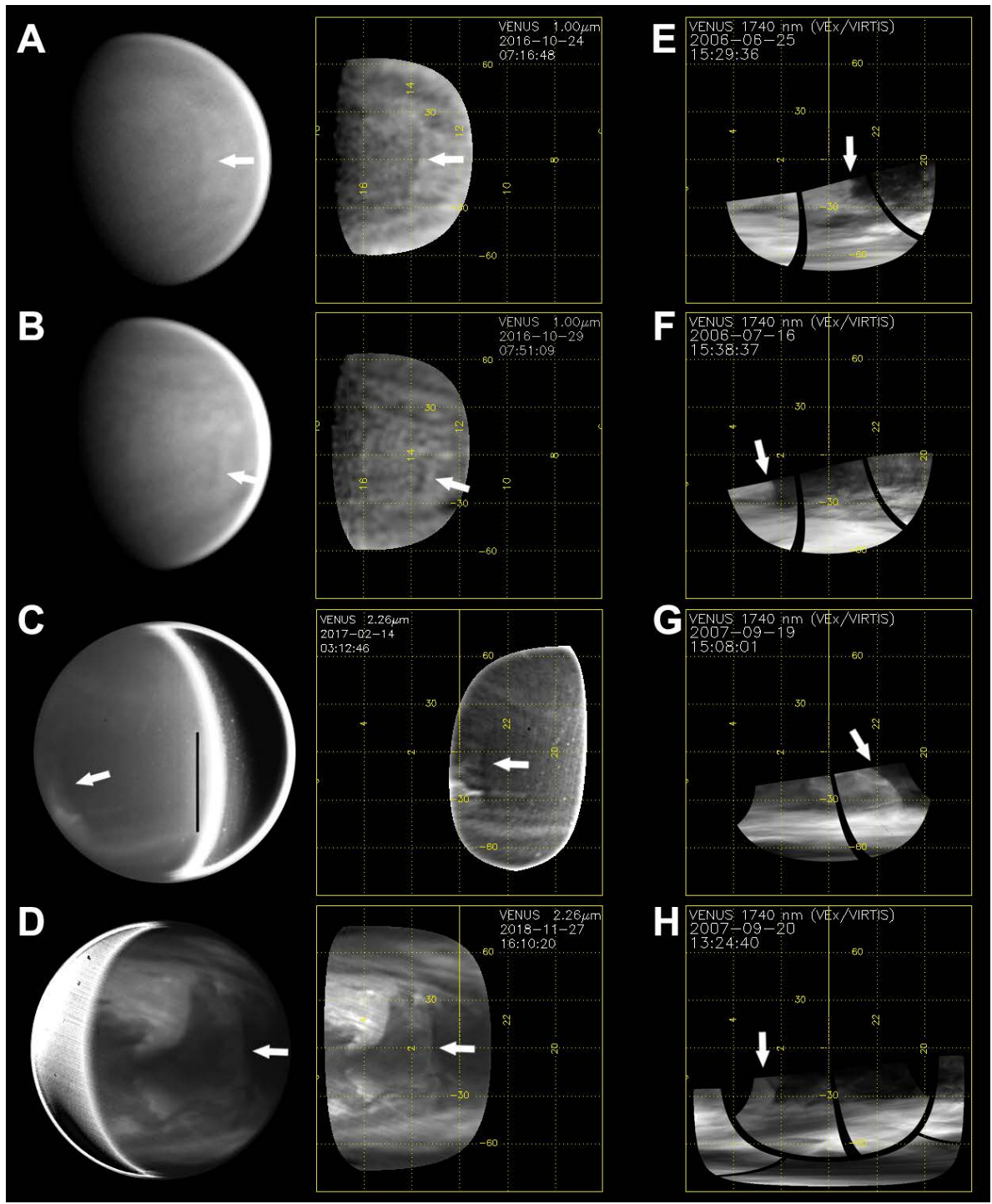

Figure 4: **The cloud disruption with ground-based observations and Venus Express.** The disruption was also identified in images of the dayside of Venus taken at 1 μm with small telescopes by observers P. Miles and A. Wesley (A–B) and on images of the nightside acquired with IRTF/SpeX using a $K_{cont}$ filter (C–D) and with VEx/VIRTIS-M at 1.74 μm (E–H). The cylindrical projections were made for the full dayside or nightside and with a spatial resolution of 0.5° per pixel.

**Supplementary Figure S5.**

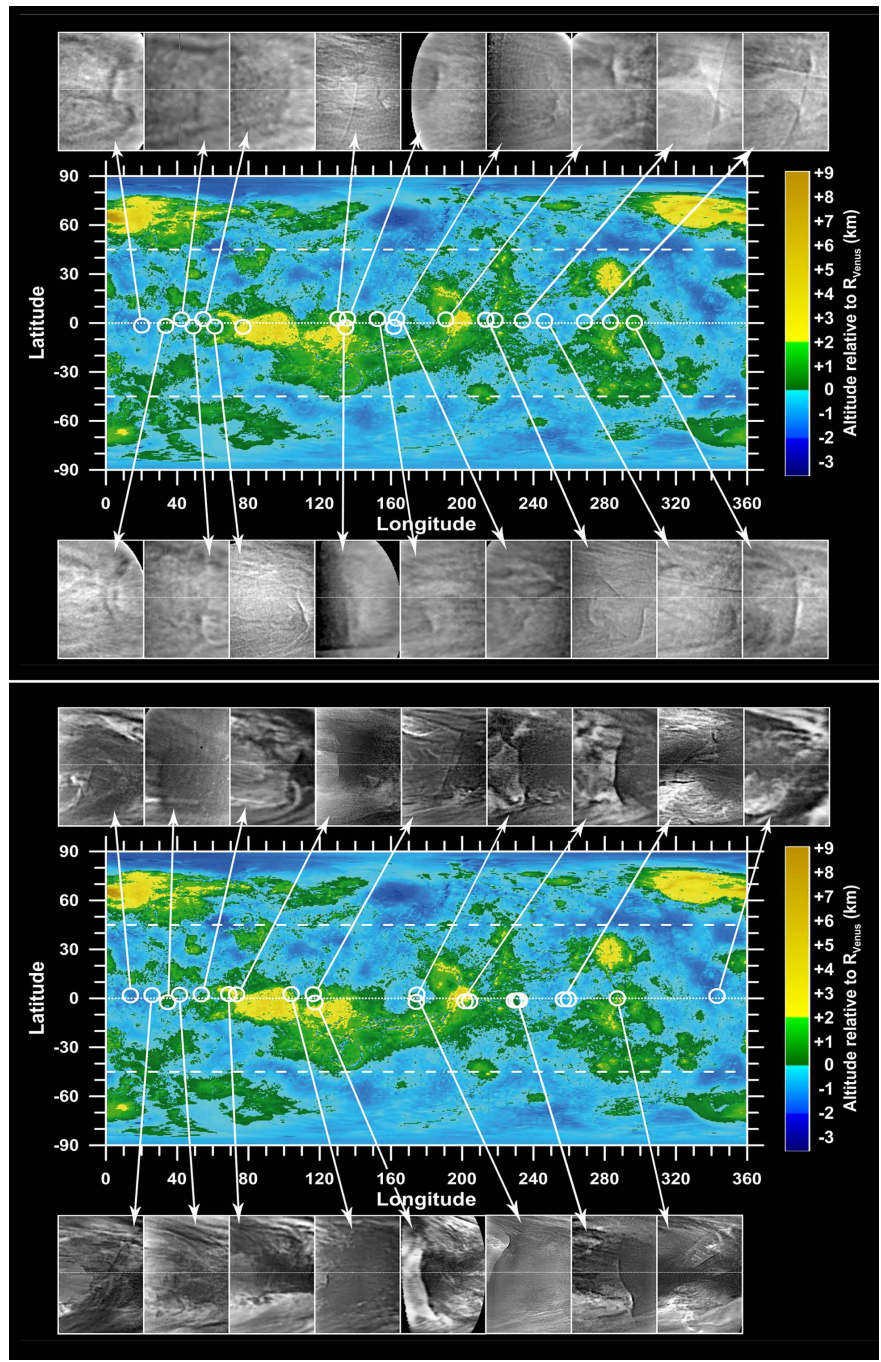

Figure 5: **Comparison of the cloud morphology and surface topography.** The morphology of the disruption in IR1/900-nm images (graph above) and IR2/2.26-$\mu$m images (graph below) is compared with the geographical location of the subsolar point during the observations (white circle). The disruptions are shown as cylindrical projections (45°N–45°S and longitude width of 60°). Latitudes 0° and 45° are marked with white dotted and dashed lines, respectively. Disruptions' morphology and hemispherical asymmetry seems unrelated to the surface elevations at the subsolar point, what might rule out the geographical location of the maximum solar heating as a probable excitation source of the disruption.

**Supplementary Figure S6.**

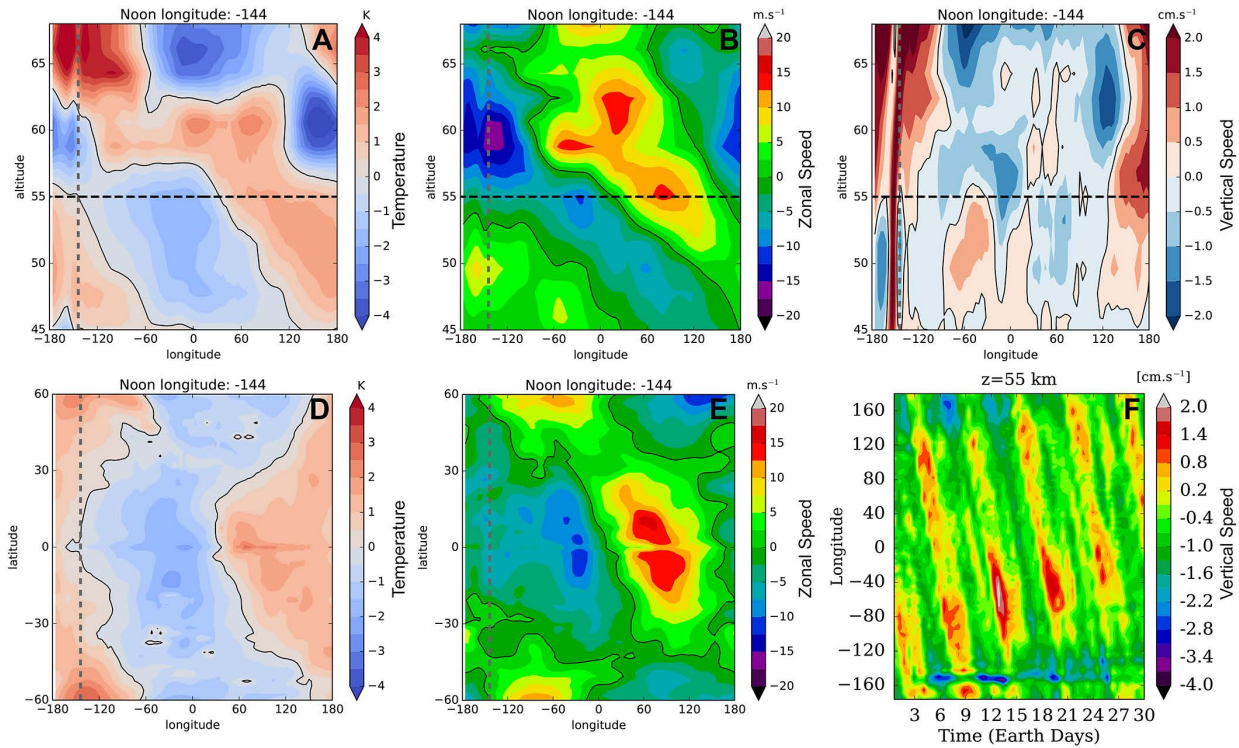

Figure 6: **Kelvin wave according to the IPSL Venus GCM: mountain wave parameterization activated.** Similarly to Figure 4 in the main article, the disturbances for the temperature, zonal and vertical speeds are shown as the vertical cross-sections within 45–68 km (A–C), while disturbances for the temperature and zonal speeds are displayed with the horizontal cross-section at 55 km height (D–E). A Hovmöller diagram for the vertical speed is also shown (F). Noon longitude and 55-km altitude level over the surface are shown with grey and black dashed lines, respectively. The mountain wave is set at longitude -160. In our simulations, the Kelvin wave is observed to pass through the mountain wave 3 Earth days later with minimal effect on the zonal wind speeds.

**Table S1.**

| Cloud level sensed | Mission/ Instrument | Wavelength ($\mu m$) | Images inspected | Time coverage | Days covered | Disruption cases |
|---|---|---|---|---|---|---|
| Night side Lower Clouds | VEx/VIRTIS | 1.74 | 4118 | from April 2006 to October 2008 | 376 | 12 |
| " | Akatsuki/IR2 | 2.26 | 238 | from March 2016 to November 2016 | 68 | 16 |
| " | IRTF/SpeX | 2.32 | 62 | from January 2017 to December 2018 | 31 | 5 |
| " | IRTF/iSHELL | 2.32 | 16 | from January 2017 to December 2018 | 8 | 2 |
| Day side Middle Clouds | Akatsuki/IR1 | 0.90 | 984 | from December 2015 to December 2016 | 173 | 19 |
| " | Small Telescope | 1.0-1.1 | 19 | from October 2016 to November 2016 | 13 | 2 |

Table 1: **Summary of imagery data set used in this work.** The dayside middle clouds are located within 50.5–56.5 km, while the nightside lower clouds are within 47.5–50.5 km (Titov et al., 2018).

**Table S2.**

| VIRTIS Cube | Date (yyyy-mm-dd) | Time (hh:mm:ss) | Wavelength ($\mu$m) | Radiance on West (W/m$^2$/sr/$\mu$m) | Radiance on East (W/m$^2$/sr/$\mu$m) | Radiance Decrease |
|---|---|---|---|---|---|---|
| VI0066_01 | 2006-06-25 | 15:29:36 | 1.74 | 0.07441 | 0.0179 | 76% |
| " | " | " | 2.26 | 0.02403 | 0.001527 | 94% |
| " | " | " | 2.32 | 0.009926 | 0.0009053 | 91% |
| VI0087_00 | 2006-07-16 | 15:10:37 | 1.74 | 0.06815 | 0.02117 | 69% |
| " | " | " | 2.26 | 0.02701 | 0.003897 | 86% |
| " | " | " | 2.32 | 0.008268 | 0.001345 | 84% |
| VI0111_04 | 2006-08-09 | 18:34:48 | 1.74 | 0.06708 | 0.00839 | 88% |
| " | " | " | 2.26 | 0.01701 | 0.0008601 | 95% |
| " | " | " | 2.32 | 0.003388 | 0.0003588 | 89% |
| VI0121_11 | 2006-08-19 | 22:19:47 | 1.74 | 0.06649 | 0.01491 | 78% |
| " | " | " | 2.26 | 0.01701 | 0.004111 | 76% |
| " | " | " | 2.32 | 0.006748 | 0.001426 | 79% |
| VI0159_00 | 2006-09-26 | 18:40:37 | 1.74 | 0.05589 | 0.01851 | 67% |
| " | " | " | 2.26 | 0.01475 | 0.001605 | 89% |
| " | " | " | 2.32 | 0.01247 | 0.0007836 | 94% |
| VI0218_06 | 2006-11-24 | 18:34:31 | 1.74 | 0.07565 | 0.01687 | 78% |
| " | " | " | 2.26 | 0.01983 | 0.001339 | 93% |
| " | " | " | 2.32 | 0.007337 | 0.000382 | 95% |
| VI0517_01 | 2007-09-19 | 15:08:01 | 1.74 | 0.1117 | 0.03523 | 69% |
| " | " | " | 2.26 | 0.04477 | 0.00683 | 85% |
| " | " | " | 2.32 | 0.01805 | 0.002687 | 85% |

Table 2: **Decrease of the Radiance across the Disruption.** The West-to-East decrease of the night side lower clouds' radiance across the disruption has been characterized for 7 cubes of the VEx/VIRTIS data set. Measurements were made for images at three wavelengths: 1.74, 2.26 and 2.32 $\mu$m.